\documentclass{iau}
\usepackage{graphicx}
\usepackage{natbib}
\usepackage{ulem}
\usepackage{amssymb}
\usepackage{times}
\usepackage[usenames,dvipsnames]{pstricks}
 \usepackage{psfrag}

\def\apj{ApJ}
\def\apjl{ApJ}
\def\apss{Ap\&SS}
\def\aap{A\&A}
\def\aapr{A\&A~Rev.}
\def\mnras{MNRAS}
\def\physrep{Phys.~Rep.}

\newcommand{\be}{\begin{equation}}
\newcommand{\ee}{\end{equation}}
\newcommand{\bary}{\begin{eqnarray}}
\newcommand{\eary}{\end{eqnarray}}
\newcommand{\en}{E_\nu}

\title[VHE neutrino expectation from FR I] 
{Very high energy neutrino expectation from Fanaroff-Riley I sources}

\author[A. Marinelli \& N. Fraija]   
{Antonio Marinelli$^1$
 \and Nissim Fraija$^2$}

\affiliation{$^1$Instituto de F\'isica, Universidad Nacional Aut\'onoma de 
M\'exico, Circuito Exterior, C.U., A. Postal 70-264, 04510 M\'exico D.F., 
M\'exico. \\ email: {\tt antonio.marinelli@pi.infn.it, antonio.marinelli@fisica.unam.mx} \\[\affilskip]
$^2$Instituto de Astronom\' ia, Universidad Nacional Aut\'onoma de M\'exico, 
Circuito Exterior, C.U., A. Postal 70-264, 04510 M\'exico D.F., M\'exico. \\email: {\tt nifraija@astro.unam.mx , Luc Binette Foundation}} 

\pubyear{2014}
\volume{313}  
\pagerange{119--126}
\jname{Extragalactic jet from every angle}
\editors{F.Massaro, C.C.Cheung, E.Lopez \& A.Siemiginowska, eds.}
\begin{document}

\maketitle

\begin{abstract}
 Fanaroff-Riley I radiogalaxies have been observed in TeV gamma-rays during the last decades. The origin of the emission processes related with this energy band is still under debate. Here we consider the case of the two closest Fanaroff-Riley I objects: Centaurus A and M87.  Their entire broadband spectral energy distributions and variability fluxes show evidences that leptonic models are not sufficient to explain their fluxes above 100 GeV.  Indeed, both objects have been imaged by LAT instrument aboard of Fermi telescope with measured spectra well connected with one-zone leptonic models. However, to explain the TeV spectra obtained with campaigns by H.E.S.S., for Centaurus A, and by  VERITAS, MAGIC and H.E.S.S. for M87, different emission processes must be introduced.  In this work we evoke hadronic scenarios to describe the TeV gamma-ray fluxes observed and to obtain the expected neutrino counterparts for each considered TeV campaign. With the obtained neutrino spectra we calculate, through Monte Carlo simulations, the expected neutrino event rate in a hypothetical Km$^{3}$ neutrino telescope and we compare the results with what has been observed by IceCube experiment up to now. 
\\
\end{abstract}

\firstsection 
\section{Introduction}
Fanaroff-Riley I (FR I) objects \citep{1985PASAu...6..130B} have been detected in the last decades at very high energy (VHE) range with gamma-ray telescopes. 
Between them Centaurus A and M87 are well known thanks to the multiple campaigns dedicated by several observatories and imaging atmospheric Cherenkov telescopes (IACTs).
Centaurus A was observed at TeV energies by the Narrabry Optical Intensity Interferometer \citep{1975ApJ...201...82G} between 1972 and 1974, by JANZOS experiment \citep{1993APh.....1..269J} 
between 1988 and 1989, by EGRET \citep{1998A&A...330...97S} between 1991 and 1995, by CANGAROO \citep{2007ApJ...668..968K} in 1999 and 2004 and by 
H.E.S.S.\citep{2009ApJ...695L..40A} with the observations performed between 2004 and 2008. On the other hand, M87 was observed at TeV gamma-ray energies by HEGRA experiment between 
1998 and 1999, and by H.E.S.S., VERITAS and MAGIC \citep{2006Sci...314.1424A,2008ApJ...679..397A,2012A&A...544A..96A}  between 2004 and 2007. The origin of the TeV 
gamma-ray emission from these radiogalaxies is still under debate. Not all the TeV gamma-ray data obtained from the IACTs are well explained within synchrotron self-Compton (SSC) scenarios 
favoring the introduction of hadronic mechanisms \citep{2012grb..confE.131F} to explain the whole spectral energy distribution (SED). The assumption of hadronic origin of the TeV gamma-ray 
spectra observed from Centaurus A and M87 gives rise to a neutrino counterpart expectation from these radiogalaxies. Here we investigate two different hadronic scenarios to describe the TeV 
gamma-ray data: the interaction of accelerated protons in the jet, with the second SSC peak, and in the giant lobes \citep{2014ApJ...783...44F}, with the thermal particle density. Therefore, we obtain 
the expected neutrino spectra considering both scenarios for M87 and Centaurus A. We introduce the obtained neutrino spectra in a Monte Carlo simulation of a hypothetical Km$^{3}$ neutrino 
telescope in the north hemisphere and we get the signal to noise ratio for one year of data-taking. Furthermore, we compare our results with the observations performed by IceCube experiment. 
Considering that no particular excess of  neutrino ($\nu_{\mu}\bar\nu_{\mu}$) track-event is expected for the obtained spectra, we estimate the time and the infrastructure needed to obtain a 
positive signal to noise ratio from these two FR I sources.
\section{Hadronic Interactions}
Radiogalaxies have been proposed as powerful accelerators of charged particles through the Fermi acceleration mechanism \citep{2007Ap&SS.309..119R}. The Fermi-accelerated protons can be described by a simple power law
\be\label{prot_esp}
\frac{dN_p}{dE_p}=A_p E_p^{-\alpha}\,,
\ee
where $\alpha$ is the power index and $A_p$ is the proportionality constant.   For this work, we consider that these protons are cooled down  by p$\gamma$ and pp interactions occurring in the jet and giant  lobes, respectively.  Both interactions produce VHE gamma-rays and neutrinos as explained in the following subsections. We hereafter use primes (unprimes) to define the quantities in a comoving (observer) frame,  c=$\hbar$=1 in natural units and redshift z$\simeq$ 0.
Charged $\pi^+$ and neutral $\pi^0$ pions are obtained from p$\gamma$ interaction through the following channels   
 \begin{eqnarray}
p\, \gamma &\longrightarrow&
\Delta^{+}\longrightarrow
\left\{
\begin{array}{lll}
p\,\pi^{0}\   &&   \mbox{fraction }2/3, \\
n\,  \pi^{+}      &&   \mbox{fraction }1/3,\nonumber
\end{array}\right. \\
\end{eqnarray}
neutral pions decay into photons, $\pi^0\rightarrow \gamma\gamma$,  carrying $20\%\,(\xi_{\pi^0}=0.2)$ of the proton's energy $E_p$.  As has been pointed out by Waxman and Bahcall  \citep{PhysRevLett.78.2292}, the photo-pion spectrum is obtained from the efficiency of this process 
\begin{equation}
f_{\pi^0,p\gamma} \simeq \frac {t'_{dyn}} {t'_{\pi^0}}  =\frac{r_d}{2\,\delta_{D}\,\gamma^2_p}\int\,d\epsilon\,\sigma_\pi(\epsilon)\,\xi_{\pi^0}\,\epsilon\int dx\, x^{-2}\, \frac{dn_\gamma}{d\epsilon_\gamma} (\epsilon_\gamma=x)\,,
\end{equation}
where $t'_{dyn}$ and $t'_{\pi^0}$ are the dynamical and the pion cooling times, $\gamma_p$ is the proton Lorentz factor, $r_{d}=\delta_{D}dt$ is the comoving dissipation radius as function of Doppler factor ($\delta_{D}$) and observational time ($t^{obs}$), $dn_\gamma/d\epsilon_\gamma$ is the spectrum of target photons,  $\sigma_\pi(\epsilon_\gamma)=\sigma_{peak}\approx 9\times\,10^{-28}$ cm$^2$ is the cross section of pion production. Solving the integrals we obtain 
{\small
\bary
f_{\pi^0,p\gamma} \simeq \frac{L_\gamma\,\sigma_{peak}\,\Delta\epsilon_{peak}\,\xi_{\pi^0}}{8\pi\,\delta_D^2\,r_d\,\epsilon_{\gamma,b}\,\epsilon_{peak}}
\cases{
\left(\frac{\epsilon_{\pi^0,\gamma,c}}{\epsilon_{0}}\right)^{-1} \left(\frac{\epsilon_{\pi^0,\gamma}}{\epsilon_{0}}\right)       &  $\epsilon_{\pi^0,\gamma} < \epsilon_{\pi^0,\gamma,c}$\cr
1                                                                                                                                                                                                            &   $\epsilon_{\pi^0,\gamma,c} < \epsilon_{\pi^0,\gamma}$\,,\cr
}
\eary
}
where $\Delta\epsilon_{peak}$=0.2 GeV,  $\epsilon_{peak}\simeq$ 0.3 GeV, $L_\gamma$ is the luminosity and $\epsilon_{\gamma,b}$ is the break energy of the seed photon field.  By considering the simple power law for a proton distribution (Eq. \ref{prot_esp}) and the conservation of the photo pion flux for this process, $f_{\pi^0,p\gamma}\,E_p\,(dN/dE)_p\,dE_p=\epsilon_{\pi^0,\gamma}\,(dN/d\epsilon)_{\pi^0,\gamma}\,d\epsilon_{\pi^0,\gamma}$, the photo-pion spectrum is given by
{\small
\bary
\label{pgammam}
\left(\epsilon^2\,\frac{dN}{d\epsilon}\right)_{\pi^0,\gamma}= A_{p\gamma,\gamma}  \cases{
\left(\frac{\epsilon_{\pi^0,\gamma,c}}{\epsilon_{0}}\right)^{-1} \left(\frac{\epsilon_{\pi^0,\gamma}}{\epsilon_{0}}\right)^{-\alpha+3}          &  $ \epsilon_{\pi^0,\gamma} < \epsilon_{\pi^0,\gamma,c}$\cr
\left(\frac{\epsilon_{\pi^0,\gamma}}{\epsilon_{0}}\right)^{-\alpha+2}                                                                                        &   $\epsilon_{\pi^0,\gamma,c} < \epsilon_{\pi^0,\gamma}$\,,\cr
}
\eary
}
\noindent 
where $\epsilon_0$ is the energy normalization; the proportionality constant of p$\gamma$ interaction is given by
\be\label{Apg}
A_{p\gamma,\gamma}= \frac{L_\gamma\,\epsilon^2_0\,\sigma_{peak}\,\Delta\epsilon_{peak}\left(\frac{2}{\xi_{\pi^0}}\right)^{1-\alpha}}{4\pi\,\delta_D^2\,r_d\,\epsilon_{\gamma,b}\,\epsilon_{peak}} \,A_p\,,
\ee
\noindent and the break photon-pion energy is given by 
 \be
\epsilon_{\pi^0,\gamma,c}\simeq 31.87\,{\rm GeV}\, \delta_D^2\, \left(\frac{\epsilon_{\gamma,b}}{ {\rm MeV}}\right)^{-1}\,.
\label{pgamma}
\ee
The Eq. (\ref{pgammam}) describes the contribution of photo-pion emission to the SED.   
\\
 Pions are also obtained from the pp interaction by means of  channel \citep{2008PhR...458..173B,2003ApJ...586...79A,2002MNRAS.332..215A}
\begin{eqnarray}
p\,+ p &\longrightarrow& \pi^++\pi^-+\pi^0 + X.
\label{pp}
\end{eqnarray}
\noindent  
Once again $\pi^0\rightarrow \gamma\gamma$, carrying in this case $33\%\,(\xi_{\pi^0}=0.33)$ of the proton's energy $E_p$. Assuming that accelerated protons interact in the lobe region, spatially constrained by $R$ and thermal particle density $n_p$, we describe the efficiency of the process through
\be
f_{\pi^0,pp}\approx R\,n_p\,k_{pp}\,\sigma_{pp}\,,
\ee
where $\sigma_{pp}\simeq 30(0.95 +0.06\,\rm{ln(E/GeV))}$ mb  is the nuclear interaction cross section and $k_{pp}=1/2$ is the inelasticity coefficient. Taking into account the proton distribution (Eq. \ref{prot_esp}) and the conservation of the photo pion flux \citep{2003ApJ...586...79A, 2012ApJ...753...40F, 2002MNRAS.332..215A}
\be\label{fpp}
f_{\pi^0 , pp}(E_p)\,E_p\,\left(\frac{dN_p}{dE_p}\right)^{obs}\,dE_p=\epsilon_{\gamma, {\pi^0}}\,\left(\frac{dN_\gamma}{d\epsilon_\gamma}\right)^{obs}_{\pi^0}\,d\epsilon_{\gamma, {\pi^0}},
\ee
then the observed gamma-ray spectrum can be written as
\begin{equation}
\label{spe_pp}
\left(\epsilon^{2}_\gamma\, \frac{dN_\gamma}{d\epsilon_\gamma}\right)^{obs}_{\pi^0}= A_{pp,\gamma}\, \left(\frac{\epsilon_{\gamma,\pi^0}}{{\rm \epsilon_0}}\right)^{2-\alpha},
\end{equation}
where the proportionality constant of pp interaction is 
\be
A_{pp,\gamma}= R\,n_p\,k_{pp}\,\sigma_{pp}\,(2/\xi_{\pi^0})^{2-\alpha}\,\epsilon_0^2\,A_p\,. 
\label{App}
\ee
The Eq. (\ref{spe_pp}) shows the contribution of pp interactions to  the spectrum of gamma rays produced in the lobes. 
%
%
\section{The VHE neutrino expectation}
The hadronic interactions described above produce a neutrino counterpart in the jet (through p$\gamma$) and in the lobes (through pp) of the AGN 
 by means of $\pi^{\pm}\rightarrow e^{\pm}+\nu_{\mu}/\bar{\nu}_{\mu}+\bar{\nu}_{\mu}/\nu_{\mu}+\nu_{e}/\bar{\nu}_{e}$. The effect of neutrino oscillations on the 
expected flux balances the number of neutrinos per flavor \citep{2008PhR...458..173B} arriving to Earth. Assuming the described interactions, we expect that the VHE gamma rays 
and the respective neutrino counterpart have a SED strictly linked to the SED of accelerated primary protons.
The spectrum of  expected neutrino can be written as:
\begin{equation}
\frac{dN_\nu}{dE_\nu}=A_\nu\,\left(\frac{E_\nu}{{\rm TeV}}\right)^{-\alpha_\nu},
\label{espneu1}
\end{equation}
where the normalization factor, A$_{\nu}$,  is calculated by correlating the neutrino flux luminosity with the TeV photon flux  \citep{2008PhR...458..173B}.  This correlation is given by:
\begin{equation}
\int \frac{dN_\nu}{dE_\nu}\,\en\,dE_\nu=K\int \frac{dN_\gamma}{dE_\gamma}\,E_\gamma\,dE_\gamma\,.
\end{equation}
Where for pp interaction should be used $K=1$ and for p$\gamma$ interaction $K=1/4$. The spectral indices for neutrino and 
gamma-ray spectrum are considered similar  $\alpha\simeq \alpha_\nu$ \citep{2008PhR...458..173B} while the carried energy is slightly different: each neutrino brings 5$\%$  of the initial proton energy ($\en=1/20\,E_p$) while  each photon brings around 16.7$\%$.  With these considerations the normalization factors are related by
\begin{equation}
A_{(pp,\nu/p\gamma,\nu)}=K\cdot A_{(pp,\gamma/p\gamma,\gamma)}\,\epsilon_0^{-2}\, (2)^{-\alpha+2},
\label{nu-gamma}
\end{equation}
where A$_{pp,\gamma}$ and A$_{p\gamma,\gamma}$ are given by the Eq. (\ref{App}) and the Eq. (\ref{Apg}) and the factor $2^{-\alpha+2}$ is introduced because the neutrino carries $1/2$ of $\gamma$ energy. 
Therefore extending the spectrum of expected neutrino to maximum energies detectable by a Km$^{3}$ Cherenkov detector array, we can obtain the number expected neutrino events detected as:
\begin{equation}
N_{ev} \approx\,T \rho_{water/ice}\,N_A\,V_{eff}\,\int_{E_{min}}^{E_{max}}\sigma_{\nu}\,A_{(pp,\nu/p\gamma,\nu)}\left(\frac{E_{\nu}}{TeV}\right)^{-\alpha}dE_{\nu}.
\label{nuMCevt}
\end{equation}
Where $N_A$ is the Avogadro number, $\rho_{water/ice}$ is the density of environment for the neutrino telescope, $E_{min}$ and $E_{max}$ are the low and high energy threshold considered,
$V_{eff}$ is the $\nu_{\mu}+\bar\nu_{\mu}$ effective volume, obtained through Monte Carlo simulation, for a hypothetical Km$^{3}$ neutrino telescope considering the neutrino source at the declination of Centaurus A and M87.
\section{Analysis and Results}
\begin{figure}[h!]
\hspace{-1.5cm}
$\begin{array}{cc}
\includegraphics[width=0.66\textwidth]{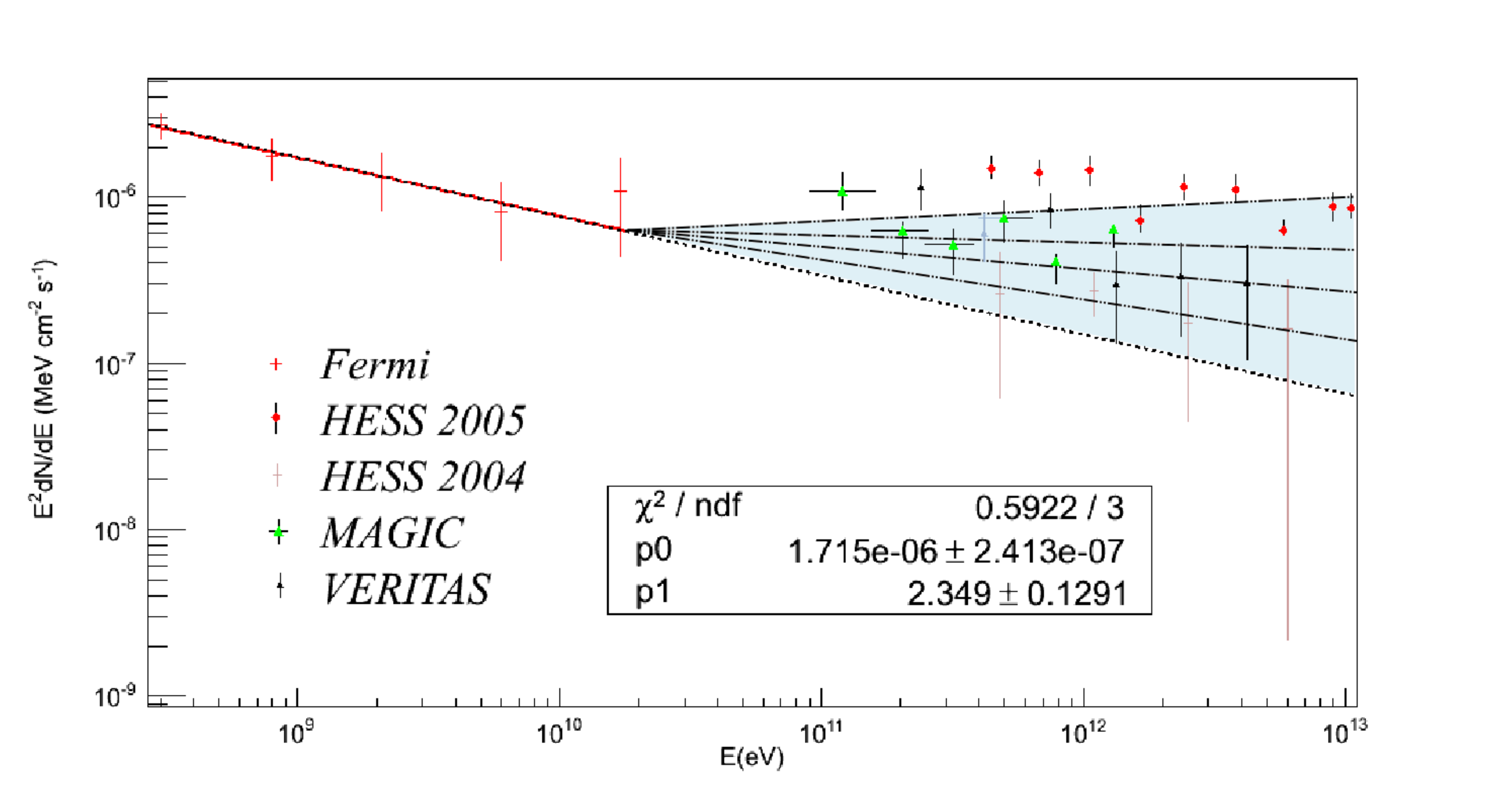} &
\includegraphics[width=0.48\textwidth]{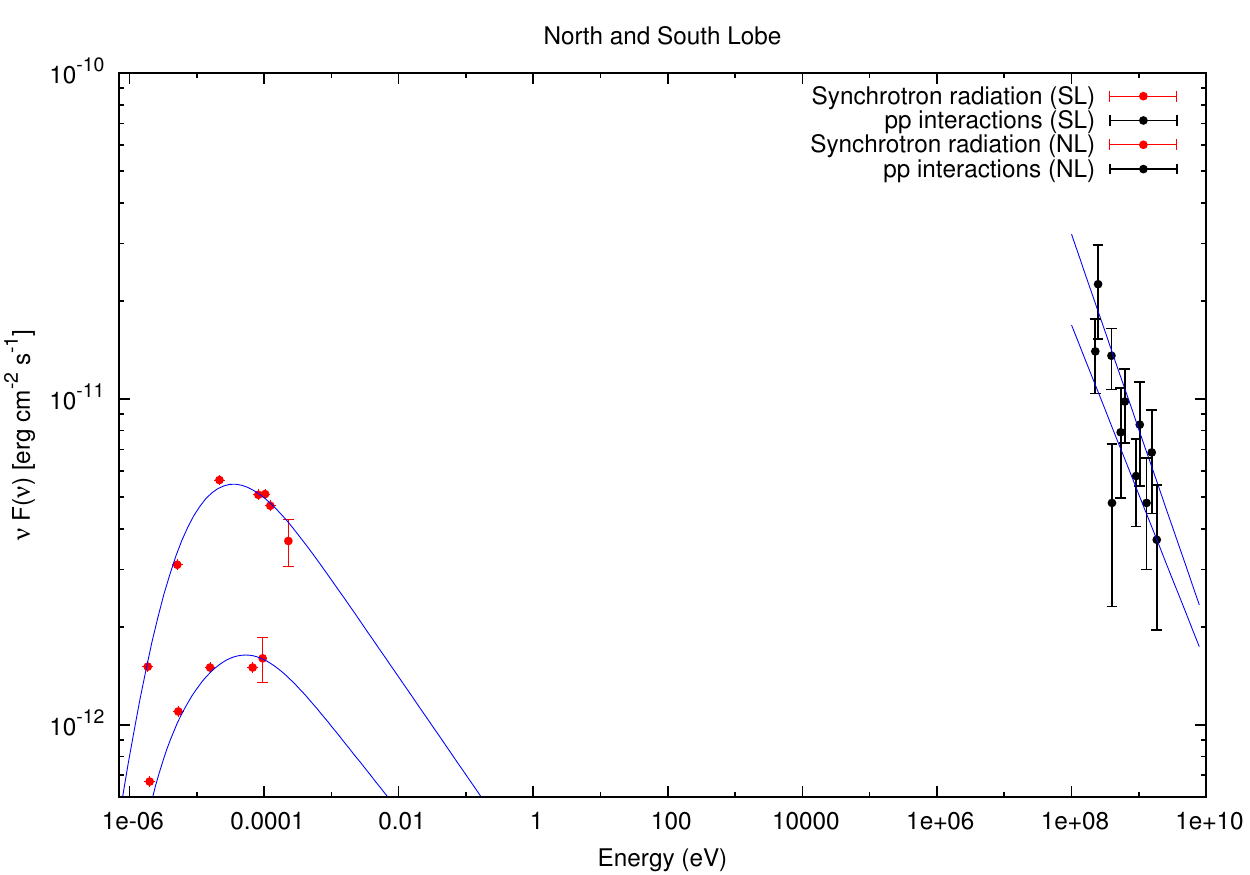}
\end{array}$
\caption{On the left: Correlations of GeV flux (Fermi data) with TeV fluxes (VERITAS, MAGIC and H.E.S.S.) for M87. On the right:
the Wilkinson Microwave Anisotropy Probe (WMAP) at the low energy of 10-5 eV and a faint gamma-ray flux imaged by the Fermi-LAT at an energy greater than 100 MeV for
Centaurus A.}\label{Spectra}
\end{figure}

\begin{figure}[h!]
\hspace{-1.5cm}
$\begin{array}{cc}
\includegraphics[width=0.6\textwidth]{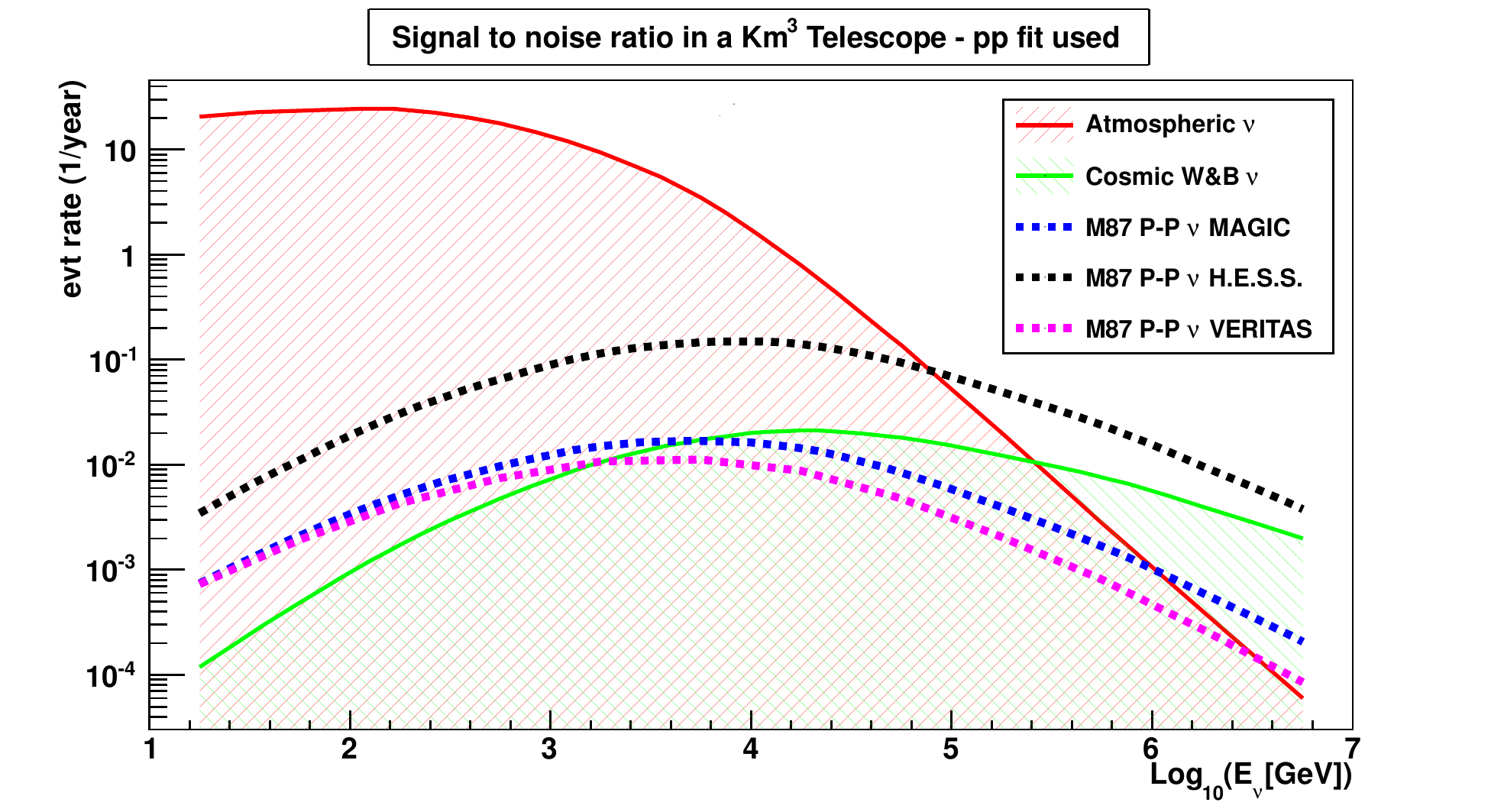} &
\includegraphics[width=0.6\textwidth]{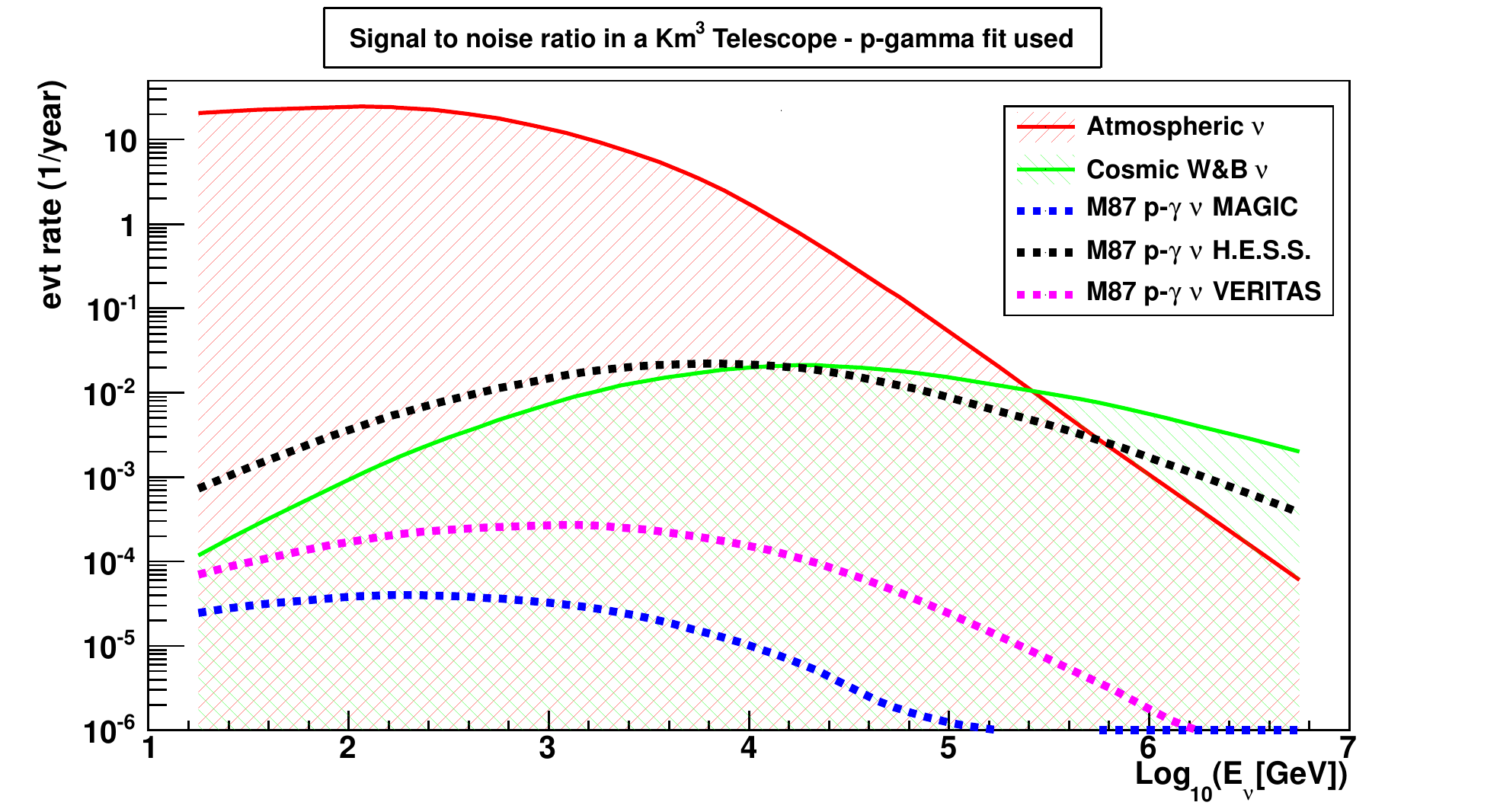} \\ 
\includegraphics[width=0.6\textwidth]{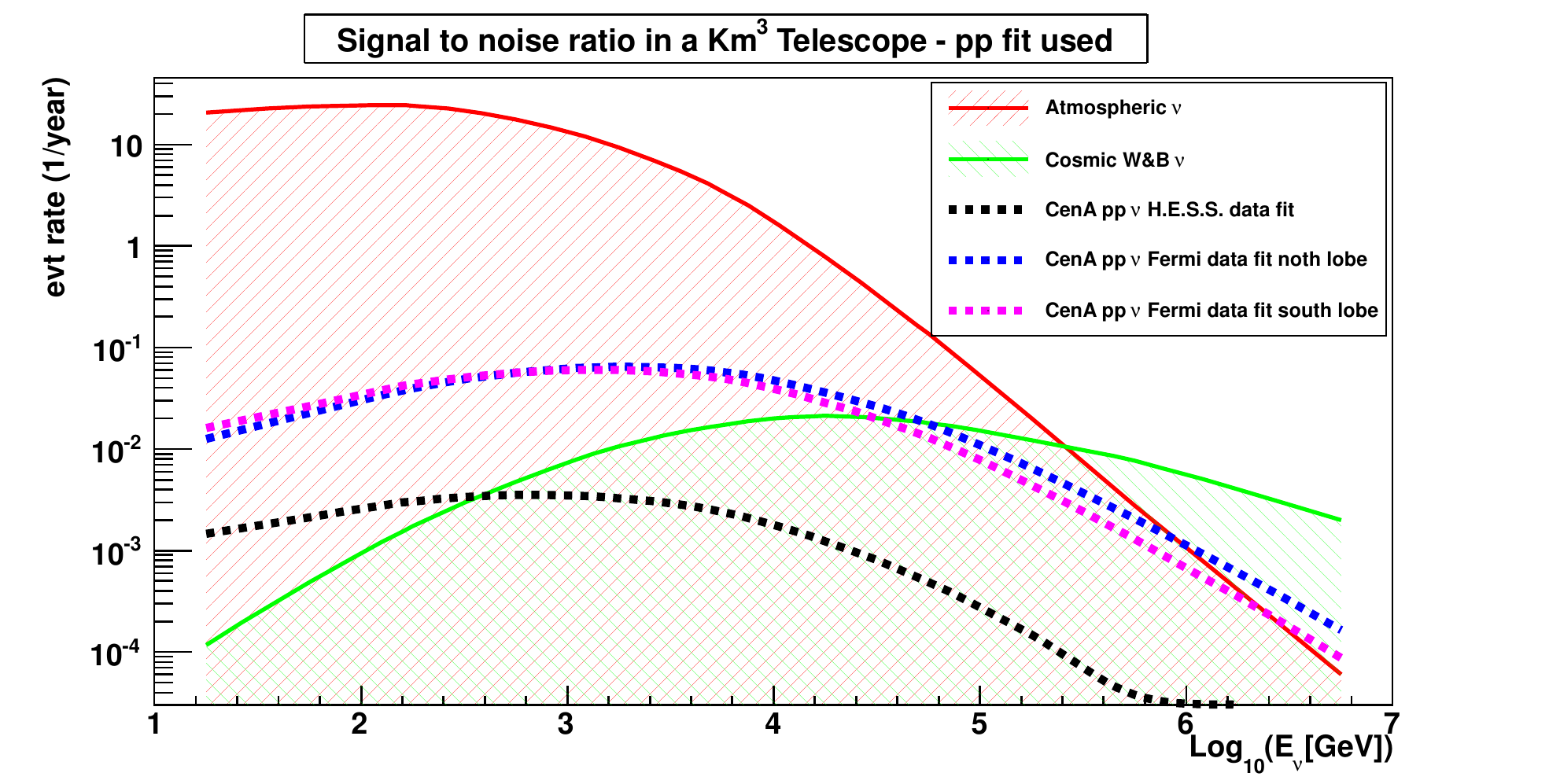} &
\includegraphics[width=0.6\textwidth]{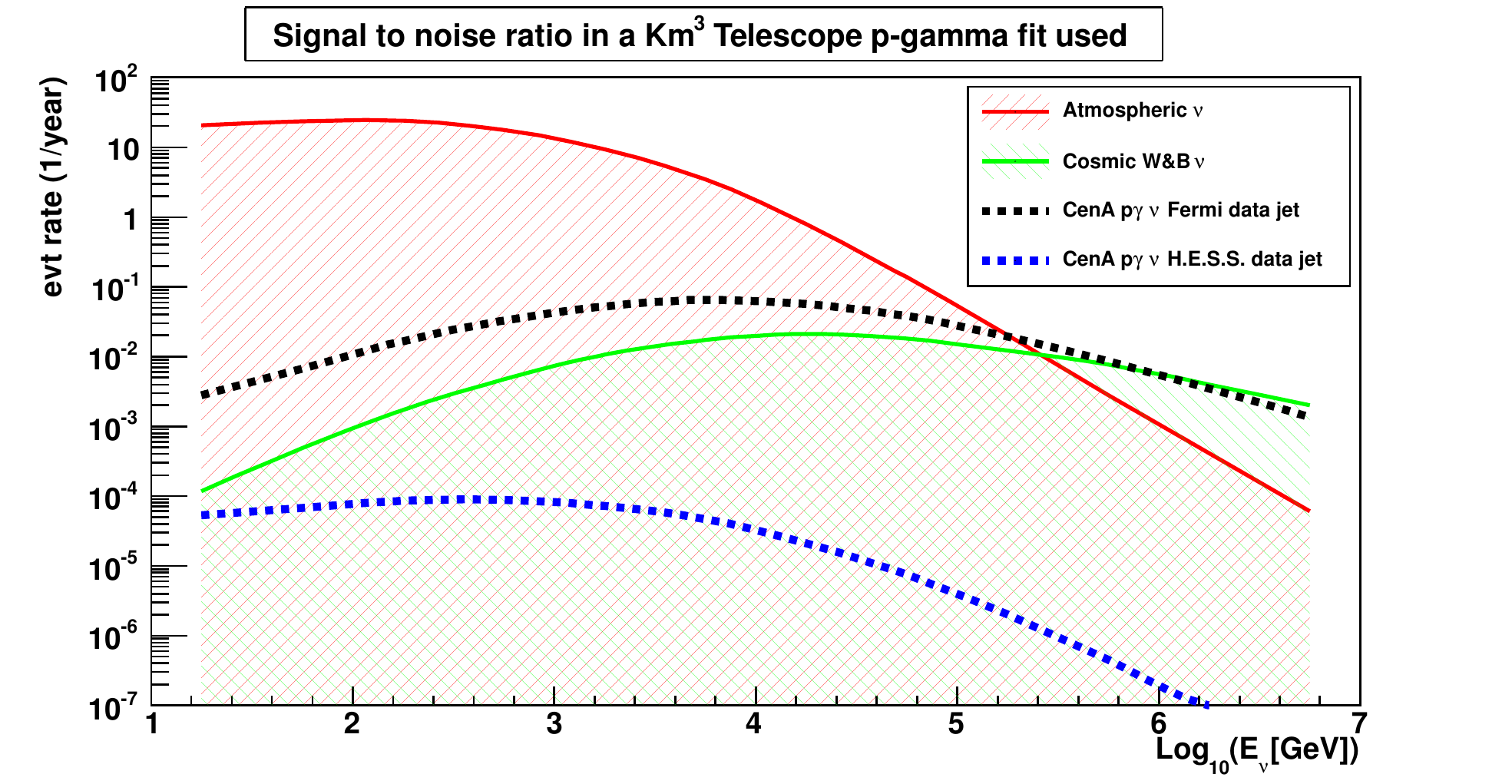}
\end{array}$
\caption{Signal to noise ratio for M87 (top) and Centaurus A (bottom) in a northern Km$^{3}$ neutrino telescope considering the region of 1$^{\circ}$ square around the positions of the two sources. As a ``background'' we take into account the atmospheric neutrinos and the extragalactic diffuse neutrinos. On the left side are showed the plots related to the pp model while on the right side are showed the plots related to the p$\gamma$ model.}\label{sig-to-noise-nu}
\end{figure}

The TeV gamma-ray fluxes collected from M87 in the last decade by H.E.S.S., VERITAS and MAGIC cannot be explained with a one-zone SSC model introduced by the Fermi collaboration \citep{2009ApJ...707...55A} to explain the SED. Fig. \ref{Spectra} shows the TeV spectra collected by IACT experiments requiring the introduction of multi-zone leptonic models or hadronic scenarios. The high-energy components in the radio galaxy Centaurus A was imaged by Fermi and H.E.S.S. observatories. Also for Centaurus A the entire SED cannot be 
described by a one-zone SSC model suggesting the hypothesis of hadronic emission components. For both Fanaroff Riley I objects we fit the TeV spectra with the hadronic models introduced in 
the Eqs. (\ref{pgammam}) and (\ref{spe_pp}), considering as a free parameters of the fits $A_{pp,\gamma}$, $A_{p\gamma,\gamma}$ and the respective $\alpha$ obtained for pp and p$\gamma$ scenarios. No extragalactic background light (EBL) models have been assumed for the spectra of Centaurus A and M87, at 3.4 Mpc \citep{1998A&ARv...8..237I} and 16.7 Mpc \citep{2010A&A...524A..71B}, respectively. For the case of Centaurus A we study the pp scenario for both giant lobes \citep{2014ApJ...783...44F} thanks to the individual SEDs obtained by Fermi satellite (see Fig. \ref{Spectra}). 
\begin{table}[ht!]
\centering\begin{tabular}{ l c c c c }
 \hline 
 \scriptsize{} & \scriptsize{Parameter} & \scriptsize{H.E.S.S.} & \scriptsize{MAGIC} & \scriptsize{VERITAS}  \\
 \hline
\scriptsize{Proportionality constant} ($10^{-13}\,{\rm TeV/cm^2/s}$) & \scriptsize{$ A_{p\gamma,\gamma} $}  &  \scriptsize{ $13.3\pm 0.096$} &  \scriptsize{ $3.38\pm 0.431$} &
\scriptsize{ $5.39\pm 0.94$}\\
\scriptsize{Power index}                    & \scriptsize{$ \alpha $}  &  \scriptsize{ $2.28\pm 0.052$} &  \scriptsize{ $2.97\pm 0.121$} & \scriptsize{ $2.70\pm 0.23$} \\
\scriptsize{Chi-square/d.o.f}                                                              & \scriptsize{$ \chi^2/{\rm d.o.f}$}  &  \scriptsize{ $14.62/7$} &  \scriptsize{ $12.59/4$} & \scriptsize{ $4.794/4$} \\
 \hline
\end{tabular}
\caption{Set of parameters obtained for the best fit of M87 SED with p$\gamma$ model}\label{table1}
\end{table}
\begin{table}[ht!]
\centering\begin{tabular}{ l c c c c}
 \hline
 \scriptsize{} & \scriptsize{Parameter} & \scriptsize{H.E.S.S.} & \scriptsize{MAGIC} & \scriptsize{VERITAS}  \\
 \hline
\scriptsize{Proportionality constant} ($10^{-13}\,{\rm TeV/cm^2/s}$) & \scriptsize{$ A_{pp,\gamma} $}  &  \scriptsize{ $12.0\pm 0.08$} &  \scriptsize{ $4.00\pm 0.43$} &
\scriptsize{ $5.11\pm 0.89$}\\
\scriptsize{Power index}                    & \scriptsize{$ \alpha $}  &  \scriptsize{ $2.22\pm 0.05$} &  \scriptsize{ $2.33\pm 0.12$} & \scriptsize{ $2.48\pm 0.20$} \\
\scriptsize{Chi-square/d.o.f.}                                                             & \scriptsize{$ \chi^2/{\rm d.o.f.}$}  & \scriptsize{ $13.93/7$} &  \scriptsize{ $7.28/4$} & \scriptsize{ $3.60/4$} \\
 \hline
\end{tabular}
\caption{Set of parameters obtained for the best fit of M87 SED with pp model}\label{table2}
\end{table}
The parameters of $A_{pp,\gamma}$, $A_{p\gamma,\gamma}$ and $\alpha$ obtained for the pp and p$\gamma$ models are reported in Tables \ref{table1} and \ref{table2} for M87, and in Tables 
\ref{table3} and \ref{table4} for Centaurus A \citep{2014MNRAS.441.1209F}. From these parameters we obtain the neutrino fluxes expected from the two scenarios, through the Eq. (\ref{nu-gamma}). 
Therefore, we calculate the neutrino ($\nu_\mu\bar{\nu}_\mu$) event rate in a Km$^{3}$ Cherenkov telescope as explained by the Eq. (\ref{nuMCevt}). We compute the $V_{eff}$ through Monte Carlo 
simulations considering the position of both sources (M87 and Centaurus A) and a Km$^{3}$ neutrino telescope located in the north hemisphere. As shown in Fig. \ref{sig-to-noise-nu}, we take into 
account also two neutrino ``backgrounds'': the atmospheric neutrinos and extragalactic diffuse neutrinos. The analysis performed did not give expectation of significant excess of M87 and Centaurus A 
neutrino signal (see Fig. \ref{sig-to-noise-nu}) in a few years of observations for the two hadronic models (pp and p$\gamma$). This result is in accordance with the observations made by IceCube 
experiment up to now. Only neutrino flux obtained with the hadronic fit (pp) of H.E.S.S. data for M87 has the possibility to be detected in several years of observation with a global neutrino network
\footnote{Global network of neutrino telescopes (IceCube, KM3NeT, ANTARES and Baikal); a infrastructure of several Km$^{3}$ and a complete coverage of the sky} (GNN). Because of the low 
equatorial declination of M87 the cross correlation between IceCube and a future northern Km$^{3}$ neutrino telescope will be favorable. 

\begin{table}[ht!]
\centering
\begin{tabular}{ l c c c}
 \hline
 Lobes (Fermi data) & \scriptsize{} & \scriptsize{North} & \scriptsize{South} \\
 \hline 
 \scriptsize{} & \scriptsize{Symbol} & \scriptsize{Value} & \scriptsize{Value} \\
 \hline
\scriptsize{Proportionality constant} ($10^{-12}\,{\rm TeV/cm^2/s}$)  & \scriptsize{$ A_{pp,\gamma} $}  &  \scriptsize{ $5.10\pm 0.96$} &  \scriptsize{ $8.07\pm 1.58$}\\
\scriptsize{Spectral index}                                                                 & \scriptsize{$\alpha$}  & \scriptsize{2.52$\pm$ 0.23} & \scriptsize{2.59$\pm$ 0.25}\\
\hline
\end{tabular}
\caption{Set of parameters obtained for the best fit of Centaurus A SED with pp model}\label{table3}
\end{table}
\begin{table}[ht!]
\centering\begin{tabular}{ l c c c}
  \hline 
 Jet & \scriptsize{} & \scriptsize{Fermi  data} & \scriptsize{H.E.S.S. data} \\
 \hline 
 \scriptsize{} & \scriptsize{Symbol} & \scriptsize{Value} & \scriptsize{Value} \\
 \hline
\scriptsize{Proportionality constant} ($10^{-12}\,{\rm TeV/cm^2/s}$)  & \scriptsize{$ A_{p\gamma,\gamma} $}  &  \scriptsize{ $2.37\pm 0.61$} &  \scriptsize{ $0.25 \pm 0.05$}\\
\scriptsize{Spectral index}                                                                 & \scriptsize{$\alpha$}  & \scriptsize{2.23$\pm$ 0.03} & \scriptsize{2.81$\pm$ 0.38}\\
\hline
\end{tabular}
\caption{Set of parameters obtained for the best fit of Centaurus A SED with p$\gamma$ model}\label{table4}
\end{table}

\section*{Acknowledgements}
This work was supported by Luc Binette scholarship and the project PAPIIT IN-108713.


\end{document}